\documentclass{JHEP3}

\usepackage{amssymb}
\usepackage{amsfonts}
\usepackage{amsbsy}
\usepackage{amsmath}
\usepackage{amsthm}
\usepackage{graphicx}
\usepackage{epstopdf}
\usepackage{multirow}
\usepackage{latexsym}
\usepackage{array}

\input cyracc.def 
\newfont{\twelvecyr}{wncyr10 at 12pt}

\def\F{\mathbb{F}}

\def\P{\mathbb{P}}

\def\n3a{t}

\def\ge{{\mathfrak{e}}}
\def\gso{{\mathfrak{so}}}
\def\gsu{{\mathfrak{su}}}

\def\gf{{\mathfrak{f}}}
\def\gg{{\mathfrak{g}}}

\newcommand{\eq}[1]{(\ref{#1})}

\title{On the Hodge structure of elliptically fibered Calabi-Yau threefolds}

\author{
Washington Taylor\\
Center for Theoretical Physics\\
Department of Physics\\
Massachusetts Institute of Technology\\
77 Massachusetts Avenue\\
Cambridge, MA 02139, USA\\
\\
{\tt wati} {\rm at} {\tt mit.edu}
}

\preprint{MIT-CTP-4370}

\abstract{The Hodge numbers of generic elliptically fibered Calabi-Yau
  threefolds over toric base surfaces fill out the ``shield''
  structure previously identified by Kreuzer and Skarke.  The
  connectivity structure of these spaces and bounds on the Hodge
  numbers are illuminated by considerations from F-theory and the
  minimal model program.  In particular, there is a rigorous bound on
  the Hodge number $h_{21}\leq 491$ for any elliptically fibered
  Calabi-Yau threefold.  The threefolds with the largest known Hodge
  numbers are associated with a sequence of blow-ups of toric bases
  beginning with the Hirzebruch surface $\F_{12}$ and ending with the
  toric base for the F-theory model with largest known gauge group.}

\begin{document}

\section{Introduction}

\begin{figure}
\begin{center}
\includegraphics[width=10cm]{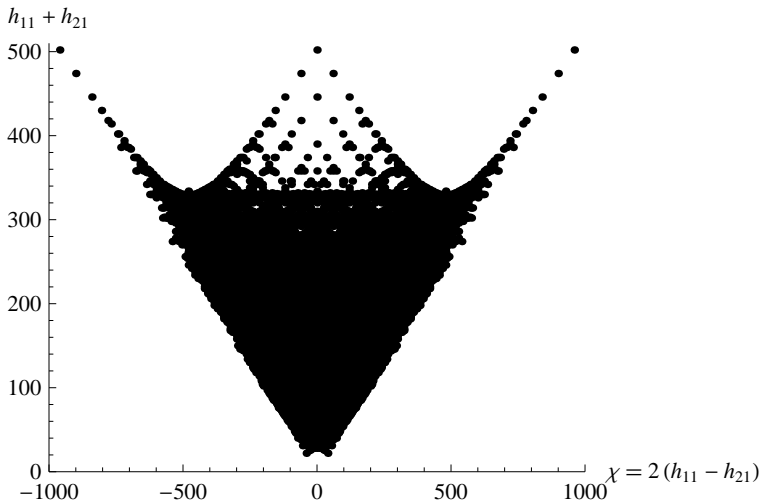}
\end{center}
\caption[x]{\footnotesize  The 30,108 distinct
Hodge numbers of the 473.8 million Calabi-Yau threefolds identified by
Kreuzer and Skarke.  Data from \cite{ks-data}.}
\label{f:ks}
\end{figure}

Since Calabi-Yau threefolds were first identified as key geometries
for superstring compactification to four dimensions \cite{chsw}, the
classification of manifolds of this type has been widely studied by
string theorists and mathematicians (See \cite{ghj} for an
introduction to the subject.)  It is still not known whether there are
a finite number of distinct topological classes of Calabi-Yau
threefolds, or if the Hodge numbers of such manifolds are bounded.
Toric geometry provides a powerful tool for describing certain classes
of Calabi-Yau manifolds.  Using a construction of Batyrev
\cite{Batyrev}, Kreuzer and Skarke \cite{Kreuzer-Skarke} produced a
comprehensive list of some 473.8 million examples of families of
Calabi-Yau manifolds associated with four-dimensional reflexive
polytopes.  These examples include manifolds with 30,108 distinct
pairs of Hodge numbers.  Graphed on a scatter plot, these Hodge
numbers take the famous ``shield'' shape (Figure~\ref{f:ks}).  While
other classes of Calabi-Yau manifolds have since been constructed (see
for example \cite{Batyrev-Kreuzer, Candelas-Davies}, \cite{Davies} for
a review), they give Hodge numbers that fit within this same general
shape.  The boundary of the set of allowed Hodge numbers has not yet
been explained in any systematic way.

Recently, D.\ Morrison and the author used an alternative approach to
construct a large class of Calabi-Yau threefolds \cite{clusters,
  mt-toric}.  Motivated by F-theory considerations, we systematically
analyzed the set of surfaces that can support elliptically fibered
Calabi-Yau threefolds.  These base surfaces are all connected in a
complicated network through transitions associated with blowing up and
down points in the surface.  From the mathematics of the minimal model
program, all such bases (aside from the trivial example of the
Enriques surface, which is connected in a more complicated way) can be
found by blowing up a series of points on $\P^2$ or the Hirzebruch
surfaces $\F_m$ \cite{Grassi, Morrison-Vafa-II}.  By analyzing the
intersection structure of irreducible effective divisors on the base
surface, in \cite{clusters} we identified specific geometric
structures (``non-Higgsable clusters'' of divisors) that characterize
the geometry of elliptically fibered Calabi-Yau threefolds.  In
particular, in \cite{mt-toric} we used this approach to construct all
smooth toric bases that support elliptic fibrations with section where
the total space is a Calabi-Yau threefold.  There are 61,539 such
toric bases.  While the approach taken in \cite{mt-toric} is also
based in toric geometry, the analysis is simplified from that of work
such as \cite{Kreuzer-Skarke} by the focus on the geometry of the
base.  The analysis of \cite{clusters} is also applicable to a
systematic analysis of non-toric bases and Calabi-Yau threefolds.

In this paper we consider the Hodge structure of generic Calabi-Yau
threefolds over the 61,539 bases constructed in \cite{mt-toric}.  We
find that the Hodge numbers $h_{11}, h_{21}$ of these manifolds are
distributed throughout the ``shield'' region identified in
\cite{Kreuzer-Skarke}.  The geometry of the base surfaces connects all
these Calabi-Yau manifolds into a connected web, in accordance with
the conjecture of Reid \cite{Reid}.  Furthermore, the structure of the
bases gives a clear geometric understanding of the outer boundary of
the shield region.  The top boundary of the region of known allowed
Hodge numbers roughly follows a specific trajectory of blow-ups of the
base $\F_{12}$ that terminates in the Calabi-Yau associated with the
F-theory model having the largest gauge group among models in this
class.  Although the analysis in this paper focuses on generic
elliptic fibrations over toric bases, a much larger class of
elliptically fibered Calabi-Yau manifolds can be realized by tuning
Weierstrass moduli, as in F-theory constructions, to produce singular
elliptic fibrations that are then resolved.  Again, though we
focus primarily here on toric bases, the methods of \cite{clusters}
suggest that similar results may hold for elliptically fibered
Calabi-Yau threefolds over non-toric bases.

In Section \ref{sec:Hodge} we describe the Hodge structure of the
generic elliptic fibrations over the bases identified in
\cite{mt-toric}.  We characterize the bounds on the region of allowed
Hodge numbers in Section \ref{sec:bounds}.  In Section
\ref{sec:extensions} we make some brief comments on extensions of this
analysis to non-generic elliptic fibrations, non-toric bases, and
non-elliptically fibered threefolds.

\section{Hodge structure}
\label{sec:Hodge}

In \cite{clusters, mt-toric} we classified the 61,539 smooth toric
compact complex surfaces that can support an elliptically fibered
Calabi-Yau threefold.  We do not repeat the analyses of those papers
here, but follow the notation of those papers and briefly summarize
some of the salient results here.  Each toric base is described by a
fan associated with a closed loop of $k$
divisors $D_1, \ldots D_k$ with
self-intersection $D_i \cdot D_i =-n_i$ and nonvanishing intersection
between adjacent divisors $D_i \cdot D_{i +1} = D_k \cdot D_1 = 1$.
The sequence of self-intersection numbers can only contain specific
subsequences (``clusters'') of self-intersections $-2$ or below.  Only
certain combinations of clusters can be connected by $-1$ curves; a
list of allowed clusters and connections appears in Table~\ref{t:connections}

The allowed toric base surfaces are all realized by blowing up a succession
of points on either $\P^2$ or $\F_m, m \leq 12$.  Some of the 61,539
bases are not strictly toric in that they arise from toric bases with
curves of self-intersection $-9, -10,$ or $-11$ that are blown up at
non-toric points to give $-12$ curves, but these bases have very similar
behavior to the true toric bases and we include these in the class of
toric bases for the purposes of this paper.
\begin{table}
\begin{center}
\begin{tabular}{| c | c  | 
c |c |
}
\hline
Cluster & gauge algebra &  $H_{\rm charged}$ & Possible connected clusters
\\
\hline
(-12) &$\ge_8$ & 0 &(-2, -2, -3) or below\\
(-8) &$\ge_7$&  0 &(-2, -3, -2) or below\\
(-7) &$\ge_7$& 28 &(-2, -3, -2) or below\\
(-6) &$\ge_6$&  0 &(-3) or below\\
(-5) &$\gf_4$&  0 &(-3, -2, -2) or below\\
(-4) &$\gso(8) $& 0 & (-4) or below\\
(-3, -2, -2)  &  $\gg_2 \oplus \gsu(2)$& 8 &any cluster\\
(-3, -2) &  $\gg_2 \oplus \gsu(2)$ &8 & (-8) or below\\
(-3)& $\gsu(3)$ & 0 &(-6) or below\\
(-2, -3, -2) &$\gsu(2) \oplus \gso(7) \oplus
\gsu(2)$&16 & (-8) or below\\
(-2, -3) &  $\gg_2 \oplus \gsu(2)$& 8 & (-5) or below\\
(-2, -2, -3)&  $\gg_2 \oplus \gsu(2)$ &8 & (-5) or below\\
(-2, -2, \ldots, -2) & no gauge group & 0 & any cluster\\
\hline
\end{tabular}
\end{center}
\caption[x]{\footnotesize Allowed clusters and connections between
  clusters by $-1$ curves in a toric surface that can be used as the
  base of an elliptic fibration.  For each cluster, the table
  indicates the resulting contribution to the gauge algebra, the
  charged matter content, and the set of clusters that can follow the
  first cluster after a $-1$ curve, where ``or below'' refers to the
  order of clusters in this table.  Note that the clusters $(-3, -2,
  -2)$ and $(-3, -2)$ are ordered; for example, a $-12$ can be connected
  by a $-1$ curve to the final $-2$ of the cluster $(-3, -2, -2)$ but
  not to the $-3$ curve.  For clarity these clusters are listed in
  both directions.}
\label{t:connections}
\end{table}

In \cite{mt-toric}, the toric bases were analyzed in the context of
F-theory \cite{Vafa-F-theory, Morrison-Vafa, Morrison-Vafa-II}
compactifications to 6 dimensions.  Though the results that we
describe here for elliptically fibered Calabi-Yau manifolds are
independent of the physics of F-theory, the tools and perspective
provided by F-theory and the minimal model program for classification
of surfaces are very helpful in illuminating the structure of these
elliptic fibrations.  In the six-dimensional supergravity theory
produced by an F-theory compactification on an elliptically fibered
Calabi-Yau threefold, gravitational anomaly cancellation relates the
numbers of tensor, vector, and hypermultiplet fields in the 6D theory
through \cite{gsw, Sagnotti, Erler} (see \cite{WT-TASI} for a review of 6D
supergravity theories, anomalies, and F-theory compactifications).
\begin{equation}
H-V= 273-29T \,.
\label{eq:anomaly}
\end{equation}
For each base, the numbers of these types of fields, as well
as the rank $r$ of the gauge group are determined from the toric data.
In particular, $T = k-3$, where $k$ is the number of curves in the fan
of the base, and the number of neutral hypermultiplets is related to
the number of Weierstrass monomials, after taking proper account of
automorphisms of the base and degrees of freedom associated with $-2$
curves not carrying a gauge group.  The gauge group and charged matter
content are determined by
the set of clusters as described in Table~\ref{t:connections}, with for example
a single $-12$ curve corresponding to an $\ge_8$ component in the
gauge algebra.  The toric data also allows for a determination of the
Hodge numbers $h_{11}, h_{21}$ of the generic elliptically fibered
threefold $X$ over each base $B$, where 
\begin{equation}
h_{11} (X) =r + T + 2 = r + k -1
\label{eq:h11}
\end{equation} and
\begin{equation}
h_{21} (X) = H_{\rm neutral} -1 = 272 + V-29T-H_{\rm charged} \,.
\label{eq:h21}
\end{equation}
in terms of the fields in the corresponding supergravity model from
F-theory.

\begin{figure}
\begin{center}
\includegraphics[width=12cm]{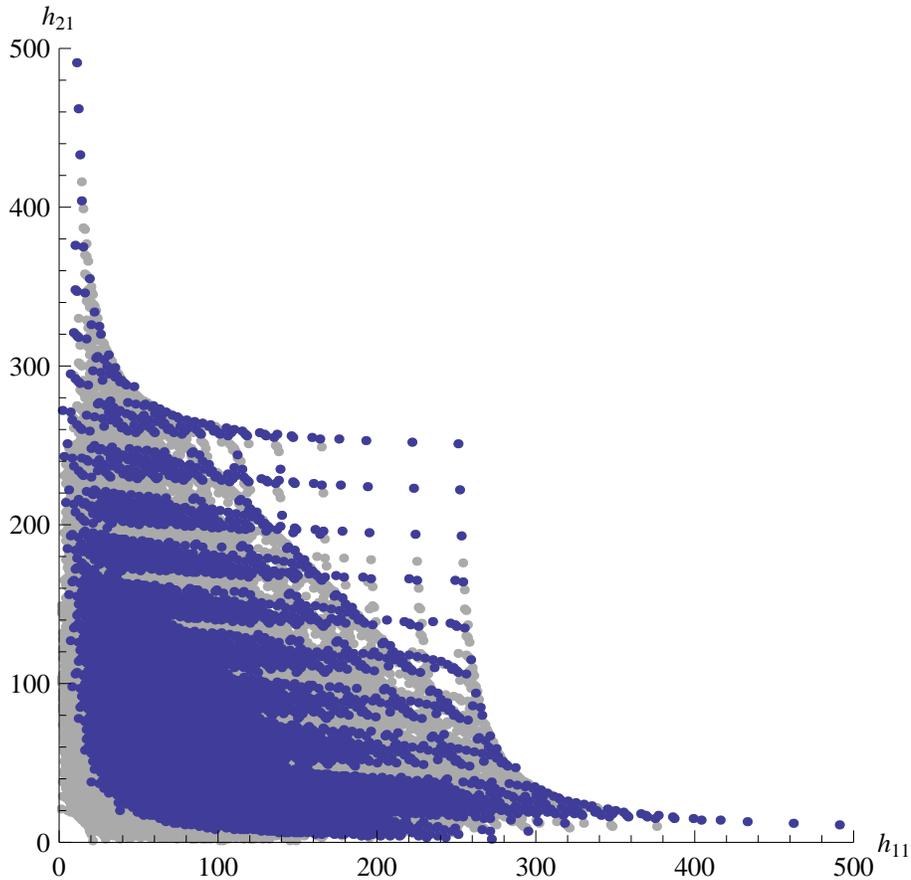}
\end{center}
\caption[x]{\footnotesize The 7524 distinct Hodge number pairs for
  generic elliptically fibered Calabi-Yau threefolds over toric bases
  (dark/blue data points).  Plot axes are Hodge numbers $h_{11},
  h_{21}$.  Kreuzer-Skarke Hodge pairs are shown in background in
  light gray for comparison.}
\label{f:Hodge}
\end{figure}

The 7524 distinct Hodge number combinations for the elliptically
fibered Calabi-Yau threefolds over the 61,539 toric bases are plotted
in Figure~\ref{f:Hodge}.  These Hodge numbers exhibit the same
``shield'' pattern seen in the Kreuzer and Skarke data.  In fact, the
Hodge data from toric bases is a proper subset of the set of Hodge
numbers from the Kreuzer and Skarke list.  It is not surprising that
the generic elliptic fibrations over toric bases all appear in the
Kreuzer and Skarke list.  What is perhaps more surprising is that the
smaller list from the set of toric bases extends through the same
general range of Hodge numbers realized on the larger list, and in
particular contains all of the larger combinations of Hodge numbers
near the upper limits of the range found by Kreuzer and Skarke.  The
list of threefolds from toric bases does not extend as close to the
origin, however, in the region of small Hodge numbers.  Note that
while the Hodge number pairs appearing in the Kreuzer and Skarke list
are invariant under mirror symmetry, which exchanges $h_{11}, h_{21}$,
the Hodge numbers for fibrations over toric bases do not have this
symmetry.  In many cases there are several distinct families of
Calabi-Yau manifolds in Kreuzer and Skarke's list that have a given
set of Hodge numbers.  The fibration structure of these manifolds can
be analyzed, for example by using the PALP software package
\cite{palp}, to identify which toric hypersurface model corresponds to
a fibration over a given toric base.

As mentioned above, all 61,539 of the Calabi-Yau threefolds
constructed in this fashion are connected through extremal transitions
associated with blowing up and down points on the base.  In F-theory
these transitions are associated with tensionless string transitions
\cite{Seiberg-Witten, Morrison-Vafa-II}.  The number of
blow-ups from $\P^2$, or one less than the number of blow-ups from
$\F_m$ corresponds to the number of tensor multiplets $T$ in the
corresponding F-theory model.  The parameter $T = k-3$ is useful in
characterizing the complexity of the base.  The models with $T = 0,1$ are
$\P^2$ and the Hirzebruch surfaces, and all lie on the left-hand side
of the diagram, ranging from $\F_{0}, \F_1,$ and $\F_2$, which all
have Hodge numbers $(h_{11}, h_{21}) = (3, 243)$, and $\P^2$ with
Hodge numbers $(2, 272)$ to $\F_{12}$ with Hodge numbers $(11, 491)$.
As more points are blown up, $T$ increases, as does the rank of the
gauge group, so $h_{11}$ monotonically increases.  At the same time,
$h_{21}$ monotonically decreases along any blow-up sequence.  The
change in $h_{21}$ denotes the number of free parameters that must be
tuned in the Weierstrass model over a given base to effect a blow-up.
Note that the monotonic increase in $h_{11}$ and decrease in $h_{21}$
is true for any sequence of blow-up operations on the base, whether or
not the base is toric.  

\section{Bounds}
\label{sec:bounds}

The shape of the upper bound on Hodge numbers in the ``shield''
configuration has been noted in previous work, but, as far as the
author of this paper knows, never explained.  From the point of view
of elliptic fibrations over toric bases, however, the upper boundary
on allowed Hodge numbers has a simple and relatively clear
interpretation.

We begin by considering the elliptically fibered Calabi-Yau threefolds
with the largest values of $h_{21}$.  Since $h_{21}$ decreases with
any blow-up of the base (toric or not), the largest values of this
parameter will appear for bases with a minimal value of $T$.  These
are the Hirzebruch surfaces $\F_m$, which (except for $\F_1$) do not
contain $-1$ curves that can be further blown down.  From \eq{eq:h21}
it is clear that the largest value of $h_{21}$ will appear when the
gauge group is largest.  This occurs for the base $\F_{12}$, where the
Hodge numbers associated with the corresponding Calabi-Yau threefold
are $(11, 491)$, since the non-Higgsable gauge algebra in the
corresponding 6D F-theory model is $\ge_8$ with $V = 248$, and there
is no charged matter.  This gives a rigorous bound $h_{21} \leq 491$
for any Calabi-Yau threefold that admits an elliptic fibration
(independent of whether the base is toric).  Indeed, $(11, 491)$ is
the Hodge number pair with the largest value of $h_{21}$ in both the
toric base and Kreuzer-Skarke lists, suggesting that this bound may
hold even outside the class of elliptically fibered Calabi-Yau
threefolds.

Going to the other corner of the allowed region, we consider
threefolds with the largest values of $h_{11}$.  As described in
\cite{mt-toric}, the largest value of $T$ arising from a threefold
associated with a toric base is $T = 193$.  The associated 6D gravity
theory has a gauge algebra containing 17 $\ge_8$ summands, 16 $\gf_4$
summands, and 32 $\gg_2 \oplus \gsu(2)$ summands, and was originally
identified in \cite{Candelas-pr, Aspinwall-Morrison-instantons}.  The
chain of self-intersections of the divisors in the toric base is
essentially 16 repeated copies of the pattern
\begin{equation}
\ldots, -12, -1, -2,  -2, -3,-1, -5, -1, -3, -2, -2, -1, -12, \ldots
\label{eq:pattern-12}
\end{equation}
with a $0$ self-intersection curve connecting to the two ends.  The
actual toric base has $-11$'s in the next-to-last positions on each
side that must be blown up at additional non-toric points, as
discussed in \cite{mt-toric}.  We denote this sequence by the shorthand
\begin{equation}
(-12//{-11}//{(-12)}^{13}//{-11}//{-12}, 0)
\label{eq:big-base}
\end{equation}
where the double slash denotes a connection between the adjacent
curves by the sequence of curves connecting the two $-12$'s in the
pattern (\ref{eq:pattern-12}), and $(-12)^{13}$ indicates 13 12's
connected by 12 copies of this pattern.  The Hodge numbers associated
with the generic threefold over this surface are $(491, 11)$, so this
Calabi-Yau manifold should be related to the generic threefold over
$\F_{12}$ by mirror symmetry, as noted in \cite{Candelas-pr}.  The
maximum value of $h_{11} =491$ is compatible with mirror symmetry and
the bound $h_{21} \leq 491$ given above for elliptically fibered
Calabi-Yau threefolds.

\begin{figure}
\begin{center}
\includegraphics[width=10cm]{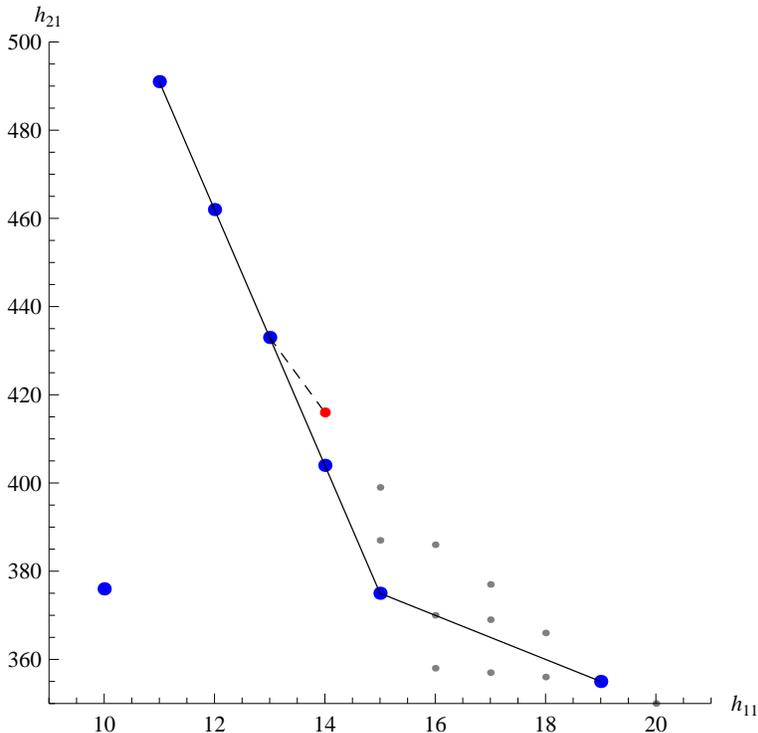}
\end{center}
\caption[x]{\footnotesize The ``tip'' of the shield region containing
  known Calabi-Yau threefold Hodge numbers with large $h_{21}$.  Large
  (blue) dots represent generic threefolds over toric bases, small
  (gray) dots are from Kreuzer-Skarke list.   Dots connected with a
  solid line represent the line of toric bases connected by blow-up
  transitions closest to the shield boundary.  Dotted line and
  mid-size (red) dot represents a threefold realized by tuning a
  Weierstrass model over the toric base corresponding to Hodge numbers
(13, 433).  (Many other Hodge pairs from the Kreuzer-Skarke list can
  be realized in a similar fashion by tuning Weierstrass moduli,
  though
only one example is depicted.)}
\label{f:tip}
\end{figure}

Now, we consider the shape of the upper boundary of the shield region,
which in Figure~\ref{f:Hodge} describes the maximum value of $h_{21}$
that can be realized as a function of $h_{11}$.  For threefolds over
toric bases, the sequence of extreme values along this curve are
determined in a simple way by the sequence of blow-ups, starting from
the base $\F_{12}$, that maximize the increase in $h_{11}$ as $h_{21}$
decreases, culminating in the base (\ref{eq:big-base}).  In the
F-theory picture, this trajectory is followed by increasing the
dimension of the gauge group (minus the number of charged matter
fields) as quickly as possible when blowing up points on the base.  In
situations not involving an increase in the gauge group or change in
matter content, a blow-up on the base trades 29 neutral hypermultiplet
moduli for one tensor multiplet, as can be seen from \eq{eq:anomaly}.
For example, blowing up one, two, three, or four toric points on the
base $\F_{12}$ cannot produce a new gauge group or matter and must
lead to changes in the Hodge numbers of $\Delta h_{11} = +1, \Delta
h_{21} = -29$.  Some of the configurations in the sequence of toric
bases realized by blowing up points on $\F_{12}$ are listed in
Table~\ref{t:connections}, along with the resulting Hodge numbers.
For the first four blow-ups, there is no way to increase the gauge
group, so $h_{21}$ drops by 29 for each blow-up and $h_{11}$ increases
by one.  The first three pairs of Hodge numbers $(11, 491), (12,
462),$ and $(13, 433)$ in this sequence have the largest values of
$h_{21}$ not only in the toric base data set but also in the Kreuzer
and Skarke data set (Figure~\ref{f:tip}).  This explains the slope of
$-29$ of the outer curve of the bounding region at the tip.  Note that
there are two distinct toric constructions in the Kreuzer and Skarke
data set with Hodge
numbers $(12, 462)$, and four with $(13, 433)$.  Using PALP, it is
easy to check that, for example, one of the $(12, 462)$ cases
corresponds to a non-generic elliptic fibration over $\F_{12}$ as
expected.  On the fifth blow-up, it is possible to produce a base
associated with a gauge group $\gg_2 \oplus \gsu(2)$ and non-Higgsable
matter.  This gives a threefold with Hodge numbers $(19, 355)$.  At
this point the slope of the bounding curve becomes less steep, as
further gauge groups can be added with additional blow-ups.

From the threefold with Hodge numbers $(13, 433)$ there is a chain of
threefolds over toric bases that roughly follows the upper boundary of
the region of allowed Hodge numbers in the Kreuzer-Skarke database
between $h_{11} = 13$ and $h_{11} =150$.  The bases associated with
these threefolds are realized by further blow-ups of $\F_{12}$ that
maximize the size of the gauge group (with matter subtracted).  Note
that in this range the threefolds with toric bases are not uniformly
at the absolute boundary of the allowed region.  The Kreuzer-Skarke
database includes some Hodge pairs that are slightly above those
realized by generic threefolds over toric bases in this region.  For example,
the Kreuzer-Skarke data includes a threefold with Hodge numbers $(14,
416)$ though no toric base gives a generic threefold with these Hodge
numbers (though $(416, 14)$ does appear in the toric base data, which
is not mirror symmetric).
We return to this example in Section~\ref{sec:tuned}
\begin{table}
\begin{center}
\begin{tabular}{| c | c  | 
c|
}
\hline
T &  curve self-intersection numbers & $h_{11}$, $h_{21}$
\\
\hline
1 & (-12, 0, 12, 0) & (11, 491)\\
2 & (-12, -1, -1, 11, 0) & (12, 462)\\
3 & (-12, -1, -2, -1,  10, 0) & (13, 433)\\
4 & (-12, -1, -2, -2, -1,  9, 0) & (14, 404)\\
5 & (-12, -1, -2, -2, -2, -1,  8, 0) & (15, 375)\\
6 & (-12, -1, -2, -2, -3, -1, -2,  8, 0) & (19, 355)\\
\hline
\vdots & \vdots & \vdots\\
\hline
60 & $(-8, -1, -2, -3, -1, -5, -1, -3, -2, -2, -1, -11//12^4, 0, 6)$ & (159, 255)\\
64 & $(9//11//12^4, 0, 6)$  & (164, 254)\\
65 & $(-8, -1, -2, -3, -2, -1, -8, -1, -2, -3, -1,
-5,$\hspace*{0.4in}&
(176, 254)\\
& \hspace*{0.7in} $-1, -3, -2, -2, -1, (-12)^5, 0, 6)$ & \\
75 & $(-10//-11//(-12)^5, 0, 6)$ & (193, 253)\\
86 & $(-11//-11//(-12)^6, 0, 6)$ & (222, 252)\\
97 & $(-12//-11//(-12)^7, 0, 6)$ & (251, 251)\\
98 & $(-12//-11//(-12)^7, -1, -1, 5)$ & (252, 222)\\
99 & $(-12//-11//(-12)^7, -1, -2, -1, 4)$ & (253, 193)\\
100 & $(-12//-11//(-12)^7, -1, -2, -2, -1, 3)$ & (254, 164)\\
\hline
\vdots & \vdots & \vdots\\
\hline
171 & $(-11//-11//(-12)^{11}//-11//-11, 0)$ & (433, 13)\\
182 & $(-12//-11//(-12)^{12}//-11//-11, 0)$ & (462, 12)\\
193 & $(-12//-11//(-12)^{13}//-11//-12, 0)$ & (491, 11)\\
\hline
\end{tabular}
\end{center}
\caption[x]{\footnotesize  Some of the toric bases that arise in the
  sequence of blow-ups from $\F_{12}$ to the maximal model with $T =
  193$, and the Hodge numbers of the generic elliptically fibered
  Calabi-Yau threefold over these bases.  This sequence of threefolds
  runs along the upper boundary of the ``shield'' region of known
  Calabi-Yau threefold Hodge numbers.}
\label{t:edge}
\end{table}

\begin{figure}
\begin{center}
\includegraphics[width=10cm]{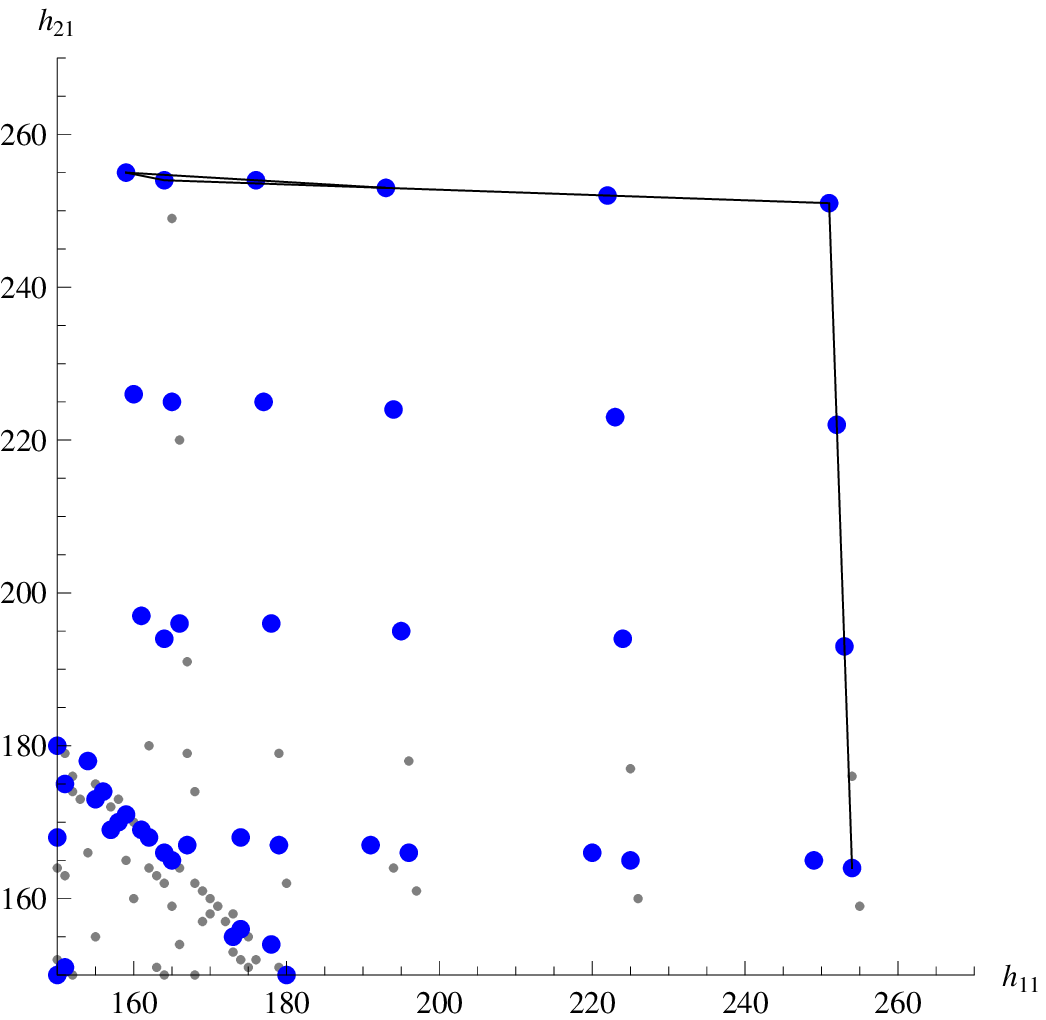}
\end{center}
\caption[x]{\footnotesize  The region around the central point of the
  ``shield'' region of known allowed Hodge numbers for Calabi-Yau
  threefolds.
Large
  (blue) dots represent generic threefolds over toric bases, small
  (gray) dots are from Kreuzer-Skarke list.   Dots connected with a
  solid line represent the line of toric bases connected by blow-up
  transitions closest to the shield boundary.}
\label{f:middle}
\end{figure}
Near the region of the central peak in the ``shield,'' the data from
generic threefolds over toric bases again contains all the Hodge pairs
realized in the Kreuzer-Skarke database (see Figure~\ref{f:middle}).
In particular, for $h_{11} > 160$ and $h_{21} > 180$, all the points
on the upper boundary of the region are realized by the sequence of
blow-ups of $\F_{12}$ mentioned above.  Part of this sequence of bases
is given in Table~\ref{t:edge}, where two separate sequences of
blow-ups connect the bases with $T = 60, 65, 75$ and $T = 60, 64, 75$.
Note that the bases in the sequence with $T = 64, 75, 86, 97$ are
connected in order by combinations of 11 blow-ups at each stage.
When, for example, the point at the intersection of the $-10$ and
following $-1$ curves is blown up in the base with $T = 75,$ the
sequence becomes $-11, -1, -2, -2, -2, -3, \ldots$.  As discussed in
\cite{clusters} this base does not support an elliptic fibration, the
point between the $-2$ and $-3$ curves leaves the Kodaira
classification and must be blown up, leading to further blow-ups.  In
the toric language this is easy to see from the dual polytope to the
toric fan, as discussed in \cite{mt-toric}; the additional blow-ups
are along curves where sections $f \in -4K, g \in -6K$ must vanish to
degrees 4, 6 once the first blow up is performed.  The ``point'' of
the central peak is associated with Hodge numbers $(251, 251)$.  The
associated toric base has self-intersections $(-12//-11//(-12)^7, 0,
6)$, and, as for $\F_{12}$, the next few blow ups cannot increase the
gauge group so decrease $h_{21}$ by 29 while increasing $h_{11}$ by 1.
It is interesting to note that these apparently very different
geometric steps --- blowing up 11 points simultaneously in the bases
with Hodge numbers $(164, 254), \ldots, ( 222, 252)$, and blowing down
single curves in the bases with Hodge numbers $(252, 222), \ldots,
(254, 164)$ --- are dual through mirror
symmetry\footnote{Though as mentioned above the toric base data does
  not necessarily give mirror symmetric Calabi-Yau constructions, the
  corresponding geometries in the Kreuzer and Skarke database can be
  identified from the Hodge numbers and polytope sizes, and are mirror
  symmetric.}.  Understanding the relationship between these
transitions may give some new insights into mirror symmetry.

Continuing down the chain, after several blow-ups the gradient again
becomes less steep, leading to a final sequence of bases carrying
elliptically fibered threefolds
with maximal
values of $h_{11}$.  Again, the sequence is the mirror image of the
initial descent with gradient $-29$, and involves 11 blow ups at each
of the final steps.

\section{Extensions}
\label{sec:extensions}

\subsection{Tuning models over toric bases}
\label{sec:tuned}

The threefolds we have considered here are the generic Weierstrass
models over each toric base.  This gives only a very small fraction of
the full set of threefolds that can be realized as elliptic fibrations
over toric bases.  As has been widely studied in the context of
F-theory, by
tuning the Weierstrass coefficients over any given base the degree of
vanishing of the discriminant over given curves can be enhanced.  This
produces a singular elliptic fibration that must be resolved to get a
smooth Calabi-Yau.  In the F-theory picture, these singularities in
the discriminant locus give rise to nonabelian gauge group factors in
the 6D supergravity theory that can be identified 
through the Kodaira classification and the
Tate algorithm \cite{Kodaira, Morrison-Vafa-II, Bershadsky-all,
Morrison-sn, Grassi-Morrison-2}.
In \cite{mt-toric} we gave an explicit description
of the basis of monomials for Weierstrass models over any of the
61,539 toric bases.  In general, from the F-theory point of view
tuning monomials to get a higher degree of vanishing of $f, g$ over a
given curve will also give rise to charged matter fields associated with the
intersection of that curve with other curves and with the rest of the
discriminant locus \cite{Bershadsky-all, Katz-Vafa, Grassi-Morrison}.

For the Hodge numbers of the smooth resolved Calabi-Yau, the effect of
tuning Weierstrass moduli to enhance the vanishing of the discriminant
locus over certain curves decreases $h_{21}$.  The appearance of gauge
groups in the corresponding F-theory model increases $h_{11}$, and the
appearance of matter decreases $h_{21}$ further.  Thus, for each of
the 61,539 generic Calabi-Yau threefolds discussed in the main part of
this paper there is a large family of additional threefolds with
larger $h_{11}$ and smaller $h_{21}$.  Anomaly cancellation conditions
in 6D supergravity strongly constrain the combinations of gauge groups
and matter content that can be realized for F-theory models associated
with elliptic fibrations over any given base \cite{universality,
  finite, KMT-II}.  In particular, the number of moduli available in
$h_{21}$ provides a bound on the complexity of the models that can be
realized over any given base through \eq{eq:h21}.  In \cite{KMT}, it
was conjectured that by tuning Weierstrass moduli any combination of
gauge groups over specific divisors and matter content compatible with
anomaly cancellation conditions could be realized in F-theory, up to
the limit imposed by the number of degrees of freedom in neutral
hypermultiplets ($h_{21}$).  We are not aware of any exceptions to
this conjecture at this time.  A systematic study of F-theory models
produced by tuning Weierstrass coefficients in this way to produce
$SU(N)$ gauge groups was carried out over Hirzebruch bases $\F_m, m =
0, 1, 2$ in \cite{KMT}, and over $\P^2$ in \cite{Morrison-Taylor}.  A
systematic analysis of the $T = 0$ 6D supergravity theories
corresponding to models over $\P^2$ with $SU(N)$ gauge groups was
carried out in \cite{0}, and the possible matter representations
compatible with anomaly cancellation were identified in this case.  In
general, there is no systematic classification of codimension two
singularities in F-theory models corresponding to different matter
representations \cite{Morrison-Taylor, Esole-Yau, Esole-Yau-II}.
Braun has systematically identified some 100,000 toric hypersurface
Calabi-Yau threefolds in the Kreuzer-Skarke database that are elliptic
fibrations over $\P^2$ \cite{Braun}.  These include many of the
$SU(N)$ models studied in \cite{0, Morrison-Taylor}, as well as a wide
range of other models.

In principle, it should be possible to systematically construct a
tremendous number of different elliptically fibered Calabi-Yau
threefolds by tuning Weierstrass moduli to achieve different F-theory
models over the 61,539 toric bases.  The Hodge numbers of these
threefolds can easily be computed through \eq{eq:h11} and \eq{eq:h21}.
It is possible that many of these threefolds will not have a
construction through the Batyrev method and will not be contained in
the Kreuzer-Skarke database.  On the other hand, because of the Hodge
number structure and reduction in $h_{21}$ when tuning moduli, these
models are likely to lie within the shield region.

A full exploration of Calabi-Yau threefolds associated with these
tuned Weierstrass models is left for future work, but we mention a few
specific examples here.  In \cite{KMT} we carried out an explicit
construction of F-theory models on $\F_m$ for small values of $m$ with
$SU(N)$ gauge groups on the divisor ${\Sigma}$ having
self-intersection ${\Sigma} \cdot {\Sigma} = -m$, and
explicitly computed the degrees of freedom used in this construction.
The simplest class of cases are $SU(N)$ models on $\F_2$.  Tuning the
Weierstrass model to realize this gauge group on ${\Sigma}$
requires fixing $N^2 -1$ moduli, and from anomaly considerations there
must be $2 N$ charged hypermultiplets in the fundamental
representation of $SU(N)$.  The Hodge numbers for the resolved
elliptically fibered threefolds in these cases are then $(2 + N,
242-N^2)$.  While most of these Hodge pairs do not appear in the
spectrum associated with toric bases, they appear in the
Kreuzer-Skarke data up to $N = 15$, which is precisely the upper limit
expected from anomaly conditions \cite{KMT}.  It is natural to expect
that these models in the Kreuzer-Skarke database are  the
resolved elliptic fibrations over $\F_2$ with enhanced gauge symmetry
in the F-theory picture, since the Hodge numbers are relatively sparse
in that region.  In fact, $(4, 238)$ (the $N = 2$ case)
is the point with $h_{11} = 4$
having the largest value of $h_{21}$, and $(5, 233)$ (the $N = 3$ case)
has the largest
$h_{21}$ for $h_{11} = 5$ aside from the $(5, 251)$ point associated
with $\F_3$.  
Direct computation with PALP confirms that, for example, the unique example with Hodge numbers $(4, 238)$ is a tuned Weierstrass model over the base $\F_2$.

Another simple class of tuned Weierstrass models with $SU(N)$ gauge
groups are those associated with elliptic fibrations over $\P^2$,
where the $SU(N)$ is realized over a degree one curve.  These models
were constructed from the supergravity and F-theory points of view in
\cite{0, Morrison-Taylor}.  The corresponding resolved Calabi-Yau
threefolds have $(h_{11}, h_{21}) = (1 + N, 271 -N (45-N)/2)$ for $N
\geq 3$, and $(3, 231)$ for $N = 2$.  These Hodge numbers appear in
the Kreuzer-Skarke database for $N$ from 2 up to the maximum expected
of 24.  These models were identified in the Kreuzer-Skarke data by
Braun \cite{Braun-private}, who has explicitly analyzed the
corresponding toric models and determined that many of them have a
fibration structure matching precisely with the resolved Calabi-Yau
manifolds associated with enhanced symmetry on $\P^2$ bases.

Finally, we can use the appearance of enhanced symmetry with tuning of
moduli to explain some of the points in the Kreuzer-Skarke database
that lie slightly above the trajectory of generic threefolds over
toric bases described in the previous section.  While in general
tuning moduli leads to a decrease in $h_{21}$ that keeps the Hodge
data below this curve, near the boundary there are cases when tuning a
small gauge group gives a larger value of $h_{21}$.  One place where
such configurations are easy to identify is in the regions where the
gradient of the shield boundary is steepest.  This occurs, for
example, near the point $(11, 491)$ associated with the $\F_{12}$
base, where the gradient is $-29$.  Tuning a gauge group on the $+12$
curve on $\F_{12}$ produces large amounts of charged matter, and
cannot give points near the boundary.  There are also no single
blow-ups of $\F_{12}$ that can be tuned to give models with enhanced
gauge symmetry and large $h_{21}$.  Blowing up $\F_{12}$ twice,
however, gives the toric base described by the sequence of curves with
self-intersections $(-12, -1, -2, -1, 10, 0)$; the generic threefold
over this base has Hodge numbers $(13, 433)$.  A gauge group $SU(2)$
on the second $-1$ curve in the F-theory picture can be shown to have
$10$ fundamental matter representations, so that the resolved
Calabi-Yau has Hodge numbers $(14, 416)$.  This is precisely the first
point observed above in the Kreuzer-Skarke data that lies above the
trajectory of maximal blow-ups of $\F_{12}$ roughly outlining the
upper boundary of the Hodge shield, as shown in  Figure~\ref{f:tip}.

All of the tuned models just described involve gauge groups on toric
divisors on the bases in the F-theory picture.  There are also more
complicated ways of tuning a gauge group.  For example, as analyzed in
\cite{0, Morrison-Taylor} for elliptic fibrations over the base
$\P^2$, gauge groups can be tuned over a divisor described by a degree
$b$ curve in the base.  As $b$ increases, the possible matter
representations become more exotic, and the complexity of the
corresponding resolved Calabi-Yau threefolds grows.  For example, for
$b = 2$ the most generic model with an $SU(N)$ gauge group has $48-4
N$ fundamental and 6 antisymmetric representations.  The Hodge numbers
for the corresponding threefolds will be $(1 + N, 271+ 2 N^2 -45 N)$.
Again, these numbers appear in the Kreuzer-Skarke database for $N$
from 3 up to (and beyond) the expected bound of 12.  Because this is a
fairly populated region, without further analysis it is unclear
whether the corresponding Calabi-Yau's are the associated elliptic
fibrations; it would be interesting to understand these and related
models further.  In general the Calabi-Yau threefolds arising from
tuning gauge groups over non-toric divisors may have no convenient
toric description, and may not appear in the Kreuzer-Skarke list.  As
another example, from anomaly analysis it seems that there should be a
6D model possible over $\P^2$ with gauge group $SU(4)$, 64 fundamental
matter fields, and one matter field in the {\bf 20} ``box''
representation \cite{0}.  No F-theory model is known for this type of
matter representation, which would correspond to an exotic codimension
two singularity on a singular divisor of degree $b = 4$.  An F-theory
model of this type would correspond to a resolved Calabi-Yau with
Hodge numbers smaller than any that appear in the Kreuzer-Skarke
database --- the smallest Hodge numbers that appear there are $(14,
14)$.  So either this model cannot be realized in F-theory, or it is a
representative of a class of Calabi-Yau threefolds not in the
Kreuzer-Skarke list.  In addition to exotic matter representations for
nonabelian gauge groups, it is also possible to tune the generic
Weierstrass model over a given base to get extra sections in the
elliptic fibration enhancing the rank of the Mordell-Weil group.  In
the F-theory picture this produces additional $U(1)$ gauge group
factors that contribute to $h_{11}$ \cite{Morrison-Vafa-II}.  Since
these factors are nonlocal, they are more difficult to describe in
F-theory (and less constrained by 6D anomalies \cite{Park-Taylor}),
though some classes of elliptic fibrations with extra sections can be
characterized systematically \cite{Klemm-lry, Esole-Yau-II, Park}.  We
have not considered elliptic fibrations with multiple sections here,
but these would provide an important direction for systematically
expanding on this work.

\subsection{Non-toric bases}

While we have focused in this paper on toric bases, a similar analysis
can be carried out for non-toric bases.  Some of the results described
above provide bounds on elliptically fibered Calabi-Yau threefolds
independent of whether the base is toric, such as the upper bound
$h_{21} \leq 491$ and the identification of the possible Hodge number
combinations for large $h_{21} > 400$.  More work would be needed,
however, beyond the analysis of Weierstrass tunings discussed in the
previous section, to produce a complete list of Hodge pairs possible
for elliptically fibered Calabi-Yau threefolds including non-toric
bases.  The analysis of \cite{clusters} places constraints on the
kinds of clusters that can appear and their connection structure even
on non-toric bases.  This leads to bounds on the gauge groups allowed
in F-theory models and constrains the set of allowed possibilities.  A
simple example of a class of non-toric bases that is explored in
\cite{mt-toric} corresponds to bases containing a number $n$ of $-4$
curves and no other curves of self-intersection $-3$ or less.  In the
F-theory context these correspond to gravity theories with gauge group
$SO(8)^n$ and no matter.  Simple constraints on possible
configurations bound $n \leq 20$, and a stronger bound closer to $n
\leq 12$ is probably possible.  Some models with $T = 9 + n$ are given
in \cite{mt-toric}, where the divisor structure on the base contains a
set of closed loops with alternating $-1, -4$ curves.  The resolved
Calabi-Yau threefolds elliptically fibered over these bases will have
$h_{11} = 11 + 5n, h_{21} = 11-n$.  These Hodge numbers appear in the
Kreuzer and Skarke list for $n = 3, \ldots, 9$.  In general, many more
non-toric bases than toric bases support elliptically fibered
Calabi-Yau threefolds, and a systematic analysis of the possibilities,
particularly for bases carrying large non-Higgsable gauge groups in
the F-theory picture, might expand the story presented here in
interesting directions.  In general, however, it seems likely that
including non-toric bases will not produce Calabi-Yau threefolds with
Hodge numbers that go significantly outside the region spanned by the
threefolds elliptically fibered over toric bases.  In particular, it
seems unlikely that non-toric bases can give elliptically fibered
threefolds with significantly larger Hodge numbers than those found
here.  The reason that it may be difficult to realize large Hodge
numbers follows similar lines to the heuristic arguments in
\cite{mt-toric} arguing that it is difficult to find a base associated
with a $T$ value much larger than $T = 193$.  The basic idea is that
the only way to get large $T$ (or large $h_{11}$) is to incorporate
many $\ge_8$ summands in the gauge algebra.  But this requires
divisors of self-intersection $-12$ in the base, and such divisors can
only be connected to clusters of the form $-2, -2, -3$.  The optimal
known base (the $T = 193$ model in Table~\ref{t:edge})
involves a linear such chain --- basically 16 copies of the
periodic sequence \eq{eq:pattern-12}.  Adding loops or extra branching
to the network of intersecting irreducible effective curves does not
seem to provide structure that could increase $T$ or $h_{11}$
significantly beyond the $T = 193$ case.  Furthermore, bases at large
$T$, like Calabi-Yau threefolds with large $h_{11} + h_{21}$, become
sparse, at least in the toric constructions known, so it is harder to
find structures that would push the bounds very far.  Thus, it seems
likely that while systematically including non-toric bases will
dramatically increase the range of possible Calabi-Yau threefold
constructions, this will not significantly modify the bounds of the
region of allowed Hodge numbers.

Note that both for the additional models found by tuning Weierstrass
parameters on toric bases discussed in the previous section, and for
threefolds fibered over non-toric bases, all these elliptically
fibered threefolds will be connected through blowing up and blowing
down points in the bases.  Thus, this very large set of threefolds are
connected in a network.

\subsection{Non-elliptically fibered threefolds}

The fact that the number of birational equivalence classes of
elliptically fibered Calabi-Yau threefolds is finite was proven  some
time ago \cite{Gross}.  A simple argument for this conclusion from the
minimal model/Weierstrass picture is given in \cite{KMT-II}.  So it is
not surprising that there are bounds to the Hodge numbers possible for
elliptically fibered Calabi-Yau threefolds.  While it will be interesting to
further analyze the precise bounds on the elliptically fibered class
of spaces, the most interesting questions involve the more general set
of Calabi-Yau threefolds without the restriction to elliptic
fibrations.  The fact that the boundary of the region of allowed Hodge
numbers seems to be the same for the sampling of toric hypersurface
models considered by Kreuzer and Skarke and for the more constrained
set of elliptic fibrations over toric bases suggests that the bounds on
Hodge numbers may be universal and apply to non-elliptically fibered
Calabi-Yau threefolds in general.

While the approach taken here does not suggest any completely general
approach to bounding the Hodge numbers for arbitrary Calabi-Yau
threefolds, it does suggest one approach which may lead to bounds at
least for the set of threefolds connected by extremal transitions.
The transitions we have described here are associated with blowing up
and down points on the base of an elliptic fibration.  These
correspond in the threefolds to transitions that connect manifolds of
different topology in more complicated ways \cite{Morrison-Vafa-II},
such as the conifold transition.  It may be enlightening to consider
the combinatorial structure of how the triple intersection product and
Mori cone of threefolds connected by blow-up transitions in the base
are related.  This may suggest a more general set of rules for
transitions that would allow for a generalization of the bounds
considered here to more general Calabi-Yau threefolds.  A related
approach has recently been used to construct novel Calabi-Yau
threefolds with small Hodge numbers using conifold-type transitions
\cite{Candelas-Davies}.

\subsection{Elliptically fibered Calabi-Yau fourfolds}

A similar exploration of elliptically fibered Calabi-Yau fourfolds
through toric 3D base manifolds may reveal structure analogous to the
``shield'' pattern for Hodge numbers of CY fourfolds.  While the set
of transitions between threefold bases is more complicated, in the
toric context the mathematics of Mori theory is well understood and it
should be possible to systematically explore the space of toric
threefold bases, as outlined briefly in \cite{mt-toric}.  Work in this
direction is currently underway.

\vspace*{0.1in}

{\bf Acknowledgements}: Particular thanks to Dave Morrison, with whom
much of the work leading up to this paper was carried out, for many
helpful discussions.
I would also like to thank Volker Braun, Thomas
Grimm, Vijay Kumar, Gabriella Martini, and Daniel Park for helpful
discussions.  This research was supported by the DOE under contract
\#DE-FC02-94ER40818.

\end{document}